\documentclass[12pt,cite,epsf,epsfig]{article}
\usepackage{epsfig}

\begin{document}

\author{S. Dev\thanks{dev5703@yahoo.com}, Shivani Gupta\thanks{shiroberts\_1980@yahoo.co.in} and Radha Raman Gautam\thanks{gautamrrg@gmail.com}}

\title{CP-odd Weak Basis Invariants for Neutrino Mass Matrices with a Texture Zero and a Vanishing Minor}
\date{\textit{Department of Physics, Himachal Pradesh University, Shimla 171005, India.}\\
\smallskip}

\maketitle
\begin{abstract}
We construct the $CP$-odd weak basis invariants in the flavor basis for all the phenomenologically viable neutrino mass matrices with a texture zero and a vanishing minor and, also, find the necessary and sufficient conditions for $CP$ invariance . We examine the interrelationships between different $CP$-odd weak basis invariants for these texture structures and investigate their implications for Dirac- and Majorana-type $CP$ violation.
\end{abstract}
\section{Introduction}
The evidence for nonvanishing neutrino masses provides a clear signal for physics beyond the Standard Model (SM). In most extensions of the SM, there can be several $CP$ violating phases. The violation of $CP$ symmetry is established in the quark sector and it is natural to expect that $CP$ violation occurs in the lepton sector too. The study of $CP$ violating phases is imperative especially in view of the development of the challenging and rather expensive experimental program to measure $CP$ violation in neutrino oscillations. In the simplest scenario of three generations, there can be one $CP$ phase in the mixing matrix in the leptonic sector. In addition if the neutrinos are Majorana particles, there can be two additional phases. It is possible to work in the parametrization in which all of three $CP$ violating phases are situated in the charged current lepton mixing matrix. Without any loss of generality, one can work in the flavor basis in which the charged lepton mass matrix is diagonal. The neutrino mass matrix in this basis will then contain all the information about $CP$ violation. The search for $CP$ violation in the leptonic sector at low energies is one of the major challenges for experimental neutrino physics. The two Majorana-type $CP$ violating phases will contribute to lepton number violating (LNV) processes like neutrinoless double beta decay while the Dirac-type $CP$ violating phase $\delta$ is expected to be measured in the experiments with superbeams and neutrino beams from neutrino factories or indirectly through the area of the unitarity triangles defined for the leptonic sector. Thus, neutrino physics provides an invaluable tool for the investigation of leptonic $CP$ violation at low energies apart from having profound implications for the physics of the early universe. It is not possible to fully reconstruct the neutrino mass matrix from the observations of feasible experiments and it is, thus, natural to employ other theoretical inputs for the reconstruction of neutrino mass matrix. Several proposals have been made in the literature to restrict the possible forms of neutrino mass matrix by reducing the number of free parameters which include the presence of texture zeros \cite{1,2,3}, hybrid textures \cite{4}, vanishing minors \cite{5} and more recently, simultaneous existence of a texture zero and a vanishing minor \cite{6}. However, not all these measures taken to reduce the number of free parameters are weak basis invariant. For example, a texture zero and a vanishing minor in a certain weak basis may not be present at all or may appear at a different place in another weak basis (WB). But two sets of leptonic mass matrices related by WB transformations contain the same physics. Thus, it is of utmost importance to analyse specific flavor models in a basis independent manner. $CP$-odd WB invariants provide invaluable tools to study $CP$ violation both in the quark and the leptonic sector. The interest in WB invariants stems from the fact that they can be evaluated and analyzed in any conveniently chosen WB and are, thus, particularly suited to the analysis of specific Ans\"{a}tze for charged leptons and neutrino mass matrices. The $CP$-odd WB invariants must vanish for $CP$ invariance to hold. Nonvanishing values of any of these WB invariants would signal $CP$ violation. Low energy $CP$-odd weak basis invariants for two texture zero neutrino mass matrices have been studied in Ref. \cite{7}. \\
The seesaw
mechanism for understanding the scale of neutrino masses is
regarded as the prime candidate not only due to its simplicity but
also due to its theoretical appeal. In the framework of type I
seesaw mechanism \cite{8} the effective Majorana mass matrix
$M_{\nu}$ is given by
\begin{equation}
 M_{\nu}= - M_D M_R^{-1} M_D^T  \nonumber \\
\end{equation}
where $M_D$ is the Dirac neutrino mass matrix and $M_R$ is the
right-handed Majorana mass matrix. It has been noted by many
authors \cite{9,10} that the zeros of the Dirac neutrino mass
matrix $M_D$ and the right-handed
 Majorana mass matrix $M_R$ are the progenitors of zeros in the effective Majorana mass matrix $M_{\nu}$. Thus,
  the analysis of zeros in $M_D$ and $M_R$ is more basic than the study of zeros in $M_{\nu}$.
However, the zeros in $M_D$ and $M_R$ may not only show as zeros in effective neutrino mass matrix. Another
 interesting possibility is that these zeros show as a vanishing minor in the effective mass matrix $M_{\nu}$. Phenomenological analysis of the case where the zeros of $M_R$ show as a vanishing minor in $M_{\nu}$ for diagonal $M_D$ has been done recently \cite{5,10}. This, however, is not the most general case.
 A more general possibility is the simultaneous existence of a texture
  zero and a vanishing minor in $M_{\nu}$.\\
 In the present work, we derive the low energy $CP$-odd WB invariants for neutrino mass matrices with a texture zero and a vanishing minor. In the presence of a texture zero and a vanishing minor, WB invariants provide the simplest tool to investigate whether a specific lepton flavor model leads to leptonic $CP$ violation at low energies. The presence of a texture zero and a vanishing minor, in general, leads to a decrease in the number of independent $CP$ violating phases. In the earlier analysis \cite{6}, it was noticed that there are correlations between the Dirac- and Majorana-type $CP$ violating phases. It is, therefore, important to examine the interrelationships between $CP$-odd WB invariants which are required to vanish as a necessary and sufficient condition for $CP$ conservation. It is the purpose of the present work to examine systematically such interrelationships in terms of the WB invariants constructed  from the elements of the neutrino mass matrix.

\section{Weak basis invariants from the neutrino mass matrix}
A texture zero and a vanishing minor of a neutrino mass matrix in a certain WB may not be present or may appear in different places in other mass matrices obtained by WB transformations so that a neutrino mass matrix with a texture zero and a vanishing minor is not WB invariant. The relevance of $CP$-odd WB invariants in the analysis of neutrino mass matrices with a texture zero and a vanishing minor is due to the fact that such neutrino mass matrices lead to a decrease in the number of independent $CP$ violating phases. A minimum number of $CP$-odd WB invariants can be found which will all vanish for $CP$ invariance to hold \cite{11,14}. A necessary and sufficient condition for low energy $CP$ invariance in the leptonic sector is that the following three WB invariants are identically zero \cite{12}:
\begin{equation}
I_1=Img Det[H_{\nu}, H_l],
\end{equation}
\begin{equation}
I_2=Img Tr[H_l M_{\nu} M_{\nu}^* M_{\nu} H_l^* M_{\nu}^*],
\end{equation}
\begin{equation}
I_3=Img Det[M_{\nu}^* H_l M_{\nu}, H_l^*].
\end{equation}
Here $M_l$ and $M_{\nu}$ are the mass matrices for the charged leptons and the neutrinos respectively and $H_l = M_l^\dagger M_l$ while $H_\nu= M_\nu^\dagger M_\nu$.\\
The invariant $I_1$ was first proposed by Jarlskog \cite{13} as a rephasing invariant measure of Dirac-type $CP$ violation. It, also, describes the $CP$ violation in the leptonic sector and is sensitive to the Dirac-type $CP$ violating phase. The invariants $I_2$ and $I_3$ which are the measures of Majorana-type $CP$ violation were first proposed by Branco, Lavoura and Rebelo \cite{14}. The invariant $I_3$ has a special feature of being sensitive to Majorana-type $CP$ violating phases even in the limit of the exactly degenerate Majorana neutrinos \cite{15}.\\
The $CP$ violation in the lepton number conserving (LNC) processes is contained in the Jarlskog $CP$ invariant $J$ which can be calculated from the WB invariant $I_1$ using the relation
\begin{equation}
I_1=-2J(m_e^2-m_\mu^2)(m_\mu^2-m_\tau^2)(m_\tau^2 - m_e^2)\times(m_1^2-m_2^2)(m_2^2-m_3^2)(m_3^2-m_1^2)
\end{equation}
if the neutrino mass matrix $M_\nu$ is a complex symmetric matrix with eigenvalues $m_1$, $m_2$ and $m_3$ and the charged lepton mass matrix $M_l$ is diagonal:
\begin{equation}
M_l=diag(m_e, m_\mu, m_\tau).
\end{equation}
Thus, we have 
\begin{equation}
I_1=-2(m_e^2-m_\mu^2)(m_\mu^2-m_\tau^2)(m_\tau^2 - m_e^2)Img(M_{ee}A_{ee}+M_{\mu\mu}A_{\mu\mu}+M_{\tau\tau}A_{\tau\tau})
\end{equation}
where the coefficients $A_{ee}$, $A_{\mu\mu}$ and $A_{\tau\tau}$ are given by
\begin{eqnarray}
A_{ee}=M_{\mu\tau}M_{e\mu}^\ast M_{e\tau}^\ast(|M_{\mu\mu}|^2-|M_{\tau\tau}|^2-|M_{e\mu}|^2+|M_{e\tau}|^2)+M_{\mu\mu}M_{e\mu}^{\ast2}\nonumber \\(|M_{e\tau}|^2-|M_{\mu\tau}|^2)+M_{\mu\mu}^\ast M_{e\tau}^{\ast2}M_{\mu\tau}^2,
\end{eqnarray}
\begin{eqnarray}
A_{\mu\mu}=M_{e\tau}M_{\mu\tau}^\ast M_{e\mu}^\ast(|M_{\tau\tau}|^2-|M_{ee}|^2-|M_{\mu\tau}|^2+|M_{e\mu}|^2)+M_{\tau\tau}M_{\mu\tau}^{\ast2}\nonumber \\(|M_{e\mu}|^2-|M_{e\tau}|^2)+M_{\tau\tau}^\ast M_{e\mu}^{\ast2}M_{e\tau}^2,
\end{eqnarray}
\begin{eqnarray}
A_{\tau\tau}=M_{e\mu}M_{e\tau}^\ast M_{\mu\tau}^\ast(|M_{ee}|^2-|M_{\mu\mu}|^2-|M_{e\tau}|^2+|M_{\mu\tau}|^2)+M_{ee}M_{e\tau}^{\ast2}\nonumber \\(|M_{\mu\tau}|^2-|M_{e\mu}|^2)+M_{ee}^\ast M_{\mu\tau}^{\ast2}M_{e\mu}^2,
\end{eqnarray}
and $M_{ij}$ ($i,j$=$e$, $\mu$ and $\tau$ ) are the elements of the neutrino mass matrix in the flavor basis.\\
Therefore, the Jarlskog $CP$ invariant measure $J$ is given by
\begin{equation}
J=\frac{Img(M_{ee}A_{ee}+M_{\mu\mu}A_{\mu\mu}+M_{\tau\tau}A_{\tau\tau})}{(m_1^2-m_2^2)(m_2^2-m_3^2)(m_3^2-m_1^2)}.
\end{equation}
This relation can be used to calculate $J$ from the mass matrices directly for any lepton mass model rotated to the WB in which $M_l$ is diagonal.\\ On the other hand, the $CP$ violation in LNV processes can be calculated from the WB invariants $I_2$ and $I_3$ given below:
\begin{eqnarray}
I_2=Img(M_{ee}M_{e\mu}^{\ast2}M_{\mu\mu}(m_e^2-m_\mu^2)^2+M_{\mu\mu}M_{\mu\tau}^{\ast2}M_{\tau\tau}(m_\mu^2-m_\tau^2)^2\nonumber\\ +M_{\tau\tau}M_{e\tau}^{\ast2}M_{ee}(m_\tau^2-m_e^2)^2+2M_{ee}M_{e\mu}^\ast M_{e\tau}^\ast M_{\mu\tau}(m_e^2-m_\mu^2)(m_e^2-m_\tau^2)+2M_{\mu\mu}M_{\mu\tau}^\ast M_{e\mu}^\ast M_{e\tau}\nonumber\\(m_\mu^2-m_\tau^2)(m_\mu^2-m_e^2)+2M_{\tau\tau}M_{e\tau}^\ast M_{\mu\tau}^\ast M_{e\mu}(m_\tau^2-m_e^2)(m_\tau^2 -m_\mu^2)) \
\end{eqnarray} 
and
\begin{eqnarray}
I_3=-2(m_e^2-m_\mu^2)(m_\mu^2- m_\tau^2)(m_\tau^2-m_e^2)\times Img(m_e^2M_{ee}B_{ee}+m_\mu^2 M_{\mu\mu}B_{\mu\mu}+m_\tau^2M_{\tau\tau}B_{\tau\tau}). \
\end{eqnarray}
where the coefficients $B_{ee}$, $B_{\mu\mu}$ and $B_{\tau\tau}$ are given by
\begin{eqnarray}
B_{ee}=M_{\mu\tau}M_{e\mu}^\ast M_{e\tau}^\ast(m_\mu^4 |M_{\mu\mu}|^2-m_\tau^4 |M_{\tau\tau}|^2-m_e^2m_\mu^2|M_{e\mu}|^2+m_e^2m_\tau^2|M_{e\tau}|^2)\nonumber\\+m_\mu^2 M_{\mu\mu}M_{e\mu}^{\ast2}(m_e^2|M_{e\tau}|^2-m_\mu^2| M_{\mu\tau}|^2)+m_\mu^2 m_\tau^2M_{\mu\mu}^\ast M_{e\tau}^{*2}M_{\mu\tau}^2,
\end{eqnarray}
\begin{eqnarray}
B_{\mu\mu}=M_{e\tau}M_{\mu\tau}^\ast M_{e\mu}^\ast(m_\tau^4 |M_{\tau\tau}|^2-m_e^4 |M_{ee}|^2-m_\mu^2 m_\tau^2|M_{\mu\tau}|^2+m_e^2m_\mu^2|M_{e\mu}|^2)\nonumber\\+m_\tau^2 M_{\tau\tau}M_{\mu\tau}^{\ast2}(m_\mu^2|M_{e\mu}|^2-m_\tau^2| M_{e\tau}|^2)+m_e^2 m_\tau^2M_{\tau\tau}^\ast M_{e\mu}^{*2}M_{e\tau}^2,
\end{eqnarray}
and
\begin{eqnarray}
B_{\tau\tau}=M_{e\mu}M_{e\tau}^\ast M_{\mu\tau}^\ast(m_e^4 |M_{ee}|^2-m_\mu^4 |M_{\mu\mu}|^2-m_e^2m_\tau^2|M_{e\tau}|^2+m_\mu^2 m_\tau^2|M_{\mu\tau}|^2)\nonumber\\+m_e^2 M_{ee}M_{e\tau}^{\ast2}(m_\tau^2|M_{\mu\tau}|^2-m_e^2| M_{e\mu}|^2)+m_e^2 m_\mu^2M_{ee}^\ast M_{\mu\tau}^{*2}M_{e\mu}^2.
\end{eqnarray}
\section{Implications for neutrino mass matrices with a texture zero and a vanishing minor}
A comprehensive phenomenological analysis of neutrino mass matrices with a texture zero and a vanishing minor has been given in Ref. \cite{6} where it was found that only six out of a total fifteen texture structures are compatible with the current data and have interesting phenomenological implications. In this section, we construct the $CP$-odd weak basis invariants for these phenomenologically allowed classes listed in Table 1. 
\begin{table}[h]
\begin{center}
\begin{small}
\begin{tabular}{|c|c|c|}
\hline  Type  & Texture Zero & Vanishing Minor \\
\hline 2A & $M_{e\mu}=0$ & $M_{\mu\mu}M_{\tau\tau}-M_{\mu\tau}^2=0$   \\
\hline 3A. &$M_{e\tau}=0$  &  $M_{\mu\mu}M_{\tau\tau}-M_{\mu\tau}^2=0$  \\
\hline 2D. &$M_{e\mu}=0$  &  $M_{ee}M_{\tau\tau}-M_{e\tau}^2=0$ \\
\hline 3F. & $M_{e\tau}=0$ &  $M_{ee}M_{\mu\mu}-M_{e\mu}^2=0$ \\
\hline 4B. &$M_{\mu\mu}=0$ & $M_{e\mu}M_{\tau\tau}-M_{\mu\tau}M_{e\tau}=0$  \\
\hline 6C. &$M_{\tau\tau}=0$  & $M_{e\mu}M_{\mu\tau}-M_{\mu\mu}M_{e\tau}=0$  \\
\hline
\end{tabular}
\end{small}
\caption{Six allowed texture structures of $M_{\nu}$ with a texture zero and a vanishing minor.}
\end{center}
\end{table}
\subsection{Class 2A and 3A}
Class 2A has a zero (1,2) element and a vanishing minor corresponding to (1,1) entry. We obtain the invariants by imposing the texture zero condition and the zero minor condition. For class $2A$, the invariants $I_1$, $I_2$ and $I_3$ are given by  
\begin{equation} 
I_1=x (|M_{\mu \mu}|^2+|M_{\mu \tau}|^2) Img(M_{ee} M_{\tau \tau} M_{e \tau}^{*2}),  
\end{equation}
\begin{equation}
I_2= (m_\tau ^2-m_e^2)^2Img(M_{ee} M_{\tau \tau} M_{e \tau}^{*2}),
\end{equation}
\begin{equation} 
I_3=x m_e^2 m_\tau ^2(m_\mu ^2|M_{\mu \mu}|^2+m_\tau ^2|M_{\mu \tau}|^2) Img(M_{ee} M_{\tau \tau} M_{e \tau}^{*2}). 
\end{equation}
where 
\begin{equation}
x=-2(m_e^2-m_\mu^2)(m_\mu^2-m_\tau^2)(m_\tau^2 - m_e^2)
\end{equation}
The WB invariants for class $3A$ can be obtained by interchanging the $\mu$ and $\tau$ indices in the above relations.
It is clear from the above equations that $CP$ violation results from the mismatch in the phases of $M_{e \tau}$, $M_{ee}$ and $M_{\tau \tau}$. A necessary and sufficient condition for the absence of $CP$ violation for class $2A$ is given by  
\begin{equation}
2arg(M_{e \tau})=arg(M_{ee}) + arg(M_{\tau \tau}).
\end{equation}
However, it should be noted that this is not the unique way to express the condition of $CP$ invariance as the condition of the vanishing minor gives us some extra freedom to express this condition in a different way which of course is equivalent to the above condition. 
The condition given above is equivalent to the condition that the neutrino mass matrix $M_\nu$ can be factorized as $PM_\nu ^rP$ where $P$ is a diagonal phase matrix diag($e^{i \phi_1}$, $e^{i \phi_2}$, $e^{i \phi_3}$) and $M_\nu ^r$ is a real matrix. Therefore, the necessary and sufficient condition for $CP$ conservation is that the neutrino mass matrix $M_\nu$ can be written as 
\begin{equation}
M_\nu = P M_\nu ^r P. 
\end{equation}
However, it is not necessary that neutrino mass matrices with a texture zero and a vanishing minor always satisfy this equation.
\subsection{Class 2D and 3F}
Neutrino mass matrices belonging to class $2D$ have a texture zero at (1,2) position and a vanishing minor corresponding to (2,2) element. For class $2D$, the invariants $I_1$, $I_2$ and $I_3$ are given by 
\begin{equation} 
I_1=-x (1+\frac{|M_{e \tau}|^2}{|M_{ee}|^2}) Img(M_{\mu \mu} M_{e e}^* M_{\mu \tau}^{*2} M_{e \tau}^2),  
\end{equation}
\begin{equation}
I_2= \frac{(m_\mu ^2 - m_\tau ^2)^2}{|M_{ee}|^2}Img(M_{\mu \mu} M_{e e}^* M_{\mu \tau}^{*2} M_{e \tau}^2),
\end{equation}
\begin{equation} 
I_3=-x (m_e^2 m_\mu ^2 m_\tau ^2 + m_\mu ^2 m_\tau ^4 \frac{|M_{e \tau}|^2}{|M_{ee}|^2}) Img(M_{\mu \mu} M_{e e}^* M_{\mu \tau}^{*2} M_{e \tau}^2). 
\end{equation}
The necessary and sufficient condition of $CP$ invariance for this class is given by
\begin{equation}
arg(M_{ee}) + 2 arg(M_{\mu \tau})=arg(M_{\mu \mu}) +2 arg(M_{e \tau}).
\end{equation}
The WB invariants for class $3F$ can be obtained by interchanging the $\mu$ and $\tau$ indices in the above relations.
\subsection{Class 4B and 6C}
The neutrino mass matrices belonging to class 4B have a texture zero at (2,2) position  and a vanishing minor corresponding to (1,2) element. The weak basis invariants for this class are given by  
\begin{equation} 
I_1=-x (1+\frac{|M_{e \mu}|^2}{|M_{\mu \tau}|^2}) Img(M_{ee} M_{\tau \tau}^* M_{e \mu }^{*2} M_{\mu \tau}^2),  
\end{equation}
\begin{equation}
I_2= (\frac{|M_{e \tau}|^2(m_\tau ^2 - m_e^2)^2}{|M_{e \mu}|^2 |M_{\mu \tau}|^2} + \frac{2 (m_e^2-m_\mu ^2) (m_e^2-m_\tau ^2)}{|M_{\mu \tau}|^2})Img(M_{ee} M_{\tau \tau}^* M_{e \mu }^{*2} M_{\mu \tau}^2),
\end{equation}
\begin{equation} 
I_3=-x (m_e^2 m_\mu ^2 m_\tau ^2 + m_\mu ^2 m_e^4 \frac{|M_{e \mu}|^2}{|M_{\mu \tau}|^2}) Img(M_{ee} M_{\tau \tau}^* M_{e \mu }^{*2} M_{\mu \tau}^2). 
\end{equation}
It is clear from these equations for $I_1$, $I_2$ and $I_3$ that the neutrino mass matrices of class $4B$ will be $CP$ invariant if the phases of the mass matrix are fine tuned to satisfy the condition
\begin{equation}
arg(M_{ee}) + 2 arg(M_{\mu \tau})=arg(M_{\tau \tau}) +2 arg(M_{e \mu}).
\end{equation}  
The weak basis invariants for class $6C$ can be obtained by interchanging the $\mu$ and $\tau$ indices in the above invariants. The conditions of $CP$ invariance for neutrino mass matrices belonging to different viable classes have been summarized in Table 2.
\begin{table}[h]
\begin{center}
\begin{small}
\begin{tabular}{|c|c|c|}
\hline  Class  & $CP$ invariance condition \\
\hline 2A & 2 $arg(M_{e \tau})$ = $arg(M_{ee})$ + $arg(M_{\tau \tau})$   \\
\hline 2D & $arg(M_{ee})$ + 2 $arg(M_{\mu \tau})$ = $arg(M_{\mu \mu})$ + 2 $arg(M_{e \tau})$ \\
\hline 4B & $arg(M_{ee})$ + 2 $arg(M_{\mu \tau})$ = $arg(M_{\tau \tau})$ + 2 $arg(M_{e \mu})$\\
\hline
\end{tabular}
\end{small}
\caption{Conditions for $CP$ invariance for some viable classes of neutrino mass matrices with a texture zero and a vanishing minor}
\end{center}
\end{table} \\
The three WB invariants $I_1$, $I_2$ and $I_3$ are related to each other for all viable classes of neutrino mass matrices. The relations are summarized in Table 3. The Dirac-type phase $\delta$ contributing to $CP$ violation in LNC processes is contained in $I_1$ while the invariants $I_2$ and $I_3$ are measures of Majorana-type $CP$ violation which contributes to LNV processes. However, the interrelationships between $I_1$, $I_2$ and $I_3$ for various viable classes of neutrino mass matrices suggest that the three $CP$ violating phases are not independent and there is only one independent physical phase in all the phenomenologically viable neutrino mass matrices with a texture zero and a vanishing minor which contributes to $CP$ violation in both LNC and LNV processes. Hence, the distinction between Dirac- and Majorana-type phases cannot be maintained in the neutrino mass matrices with a texture zero and a vanishing minor.
\begin{table}[h]
\begin{large}
\begin{center}
\begin{tabular}{|c|c|c|}
\hline  Class  & $\frac{I_1}{I_2}$ & $\frac{I_1}{I_3}$ \\
\hline 2A & $\frac{2(m_e^2-m_\mu ^2)(m_\mu ^2-m_\tau ^2)(|M_{\mu \mu}|^2+|M_{\mu \tau}|^2)}{(m_e ^2-m_\tau ^2)}$ & $\frac{(|M_{\mu \mu}|^2+|M_{\mu \tau}|^2)}{(m_e^2 m_\mu ^2 m_\tau ^2 |M_{\mu \mu}|^2 + m_e^2 m_\tau ^4 |M_{\mu \tau}|^2)}$  \\
\hline 3A &$\frac{2(m_e ^2-m_\tau^2)(m_\mu ^2-m_\tau ^2)(|M_{\tau \tau}|^2+|M_{\mu \tau}|^2)}{(m_\mu ^2-m_e ^2)}$& $\frac{(|M_{\tau \tau}|^2+|M_{\mu \tau}|^2)}{(m_e^2 m_\mu ^2 m_\tau ^2 |M_{\tau \tau}|^2 + m_e^2 m_\mu ^4 |M_{\mu \tau}|^2)}$   \\
\hline 2D &$\frac{2(m_e^2-m_\mu ^2)(m_\tau ^2-m_e^2)(|M_{ee}|^2+|M_{e \tau}|^2)}{(m_\mu ^2-m_\tau ^2)}$& $\frac{(|M_{ee}|^2+|M_{e \tau}|^2)}{(m_e^2 m_\mu ^2 m_\tau ^2 |M_{ee}|^2 + m_\mu ^2 m_\tau ^4 |M_{e \tau}|^2)}$   \\
\hline 3F & $\frac{2(m_e^2-m_\mu ^2)(m_\tau ^2-m_e ^2)(|M_{ee}|^2+|M_{e \mu}|^2)}{(m_\tau ^2-m_\mu ^2)}$& $\frac{(|M_{ee}|^2+|M_{e \mu}|^2)}{(m_e^2 m_\mu ^2 m_\tau ^2 |M_{ee}|^2 + m_\tau ^2 m_\mu ^4 |M_{e \mu}|^2)}$   \\
\hline 4B &$\frac{2(m_e^2-m_\mu ^2)(m_\mu ^2-m_\tau ^2)(|M_{e \mu}|^2+|M_{\mu \tau}|^2)}{(\frac{|M_{e \tau}|^2 (m_\tau ^2-m_e^2)}{|M_{e \mu}|^2}+2 (m_\mu ^2-m_e^2))}$& $\frac{(|M_{e \mu}|^2+|M_{\mu \tau}|^2)}{(m_e^2 m_\mu ^2 m_\tau ^2 |M_{\mu \tau}|^2 + m_\mu ^2 m_e^4 |M_{e \mu}|^2)}$   \\
\hline 6C &$\frac{2(m_e ^2-m_\tau ^2)(m_\mu ^2-m_\tau ^2)(|M_{e \tau}|^2+|M_{\mu \tau}|^2)}{(\frac{|M_{e \mu}|^2 (m_e ^2-m_\mu ^2)}{|M_{e \tau}|^2}+2 (m_e^2-m_\tau ^2))}$& $\frac{(|M_{e \tau}|^2+|M_{\mu \tau}|^2)}{(m_e^2 m_\mu ^2 m_\tau ^2 |M_{\mu \tau}|^2 + m_\tau ^2 m_e^4 |M_{e \tau}|^2)}$  \\
\hline
\end{tabular}
\caption{Ratios $\frac{I_1}{I_2}$ and $\frac{I_1}{I_3}$ for all viable classes of neutrino mass matrices with a texture zero and a vanishing minor. }
\end{center}
\end{large}
\end{table}
\section{Conclusions}
We have calculated the $CP$-odd weak basis invariants for LNC and LNV processes for all the phenomenologically viable neutrino mass matrices with a texture zero and a vanishing minor in the flavor basis. All these invariants must vanish for $CP$ to be conserved. We find that neutrino mass matrices with a texture zero and a vanishing minor are, in general, $CP$ violating unless their phases are fine tuned. The three WB invariants in each class are found to be related to each other so that all viable classes of neutrino mass matrices with a texture zero and a vanishing minor have only one independent physical phase which contributes to $CP$ violation in both LNC and LNV processes, and, hence, cannot be labelled as either Dirac or Majorana.

\textbf{\textit{\Large{Acknowledgements}}}

The research work of S. D. is supported by the University Grants
Commission, Government of India \textit{vide} Grant No. 34-32/2008
(SR).  S. G. and R. R. G. acknowledge the financial support provided by the Council for Scientific and Industrial Research (CSIR), Government of India.

\end{document}